# All-Electrical Spin Field Effect Transistor in van der Waals Heterostructures at Room Temperature

André Dankert*, Saroj P. Dash[†]

*Department of Microtechnology and Nanoscience, Chalmers University of Technology, SE-41296, Göteborg, Sweden.*

**Spintronics aims to exploit the spin degree of freedom in solid state devices for data storage and information processing technologies.[1,2] The fundamental spintronic device concepts such as creation, manipulation and detection of spin polarization has been demonstrated in semiconductors[3,4,5] and spin transistor structures[6,7,8,9] using both the electrical and optical methods. However, an unsolved challenge in the field is the realization of all electrical methods to control the spin polarization and spin transistor operation at ambient temperature.[2] For this purpose, two-dimensional (2D) crystals offer a unique platform due to their remarkable and contrasting spintronic properties, such as weak spin-orbit coupling (SOC) in graphene and strong SOC in molybdenum disulfide ($MoS_2$).[10,11] Here we combine graphene and $MoS_2$ in a van der Waals heterostructure to realize the electric control of the spin polarization and spin lifetime, and demonstrated a spin field-effect transistor (spin-FET) at room temperature in a non-local measurement geometry. We observe electrical gate control of the spin valve signal due to pure spin transport and Hanle spin precession signals in the graphene channel in proximity with $MoS_2$ at room temperature. We show that this unprecedented control over the spin polarization and lifetime stems from the gate-tuning of the Schottky barrier at the $MoS_2$/graphene interface and $MoS_2$ channel conductivity leading to spin interaction with high SOC material. The all-electrical creation, transport and control of the spin polarization in a spin-FET device at room temperature is a substantial step in the field of spintronics. It opens a new platform for the interplay of spin, charge and orbital degrees of freedom for testing a plethora of exotic physical phenomena,[12] which can be key building blocks in future device architectures.**

*andre.dankert@chalmers.se; [†]saroj.dash@chalmers.se

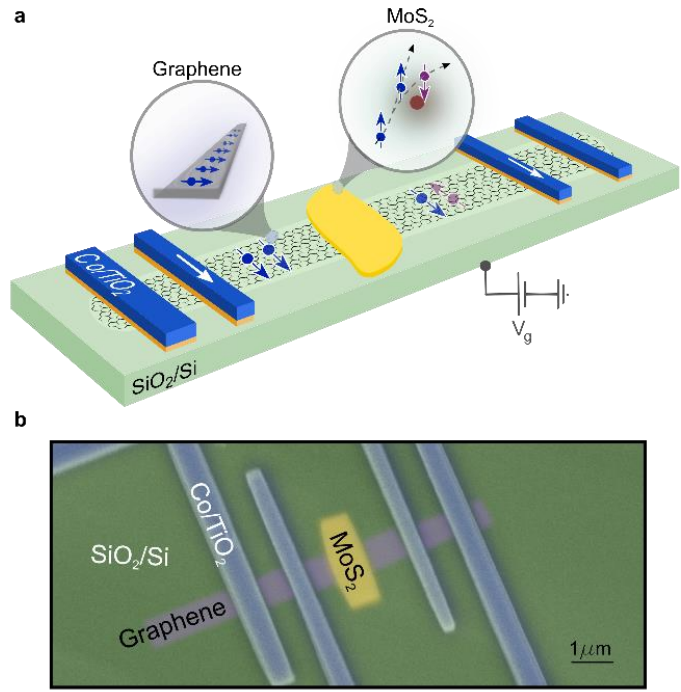

**Figure 1: MoS$_2$/Graphene van der Waals heterostructure spin-FET: a.** Schematics of graphene/MoS$_2$ heterostructure channel with ferromagnetic source and drain contacts. This structure allows spin injection into graphene from the source, diffusive spin transport in the graphene/MoS$_2$ channel, spin manipulation by a gate voltage and detection of spin signal by the drain. **b.** Colored scanning electron microscope image of a fabricated spin-FET with a CVD-graphene/MoS$_2$ heterostructure channel and multiple ferromagnetic tunnel contacts of TiO$_2$(1nm)/Co(80 nm). The devices are fabricated on Si/SiO$_2$ substrate, which is used as a back gate electrode for control of the spin polarization in the channel.

Two-dimensional (2D) crystals offer a unique platform for spintronics due to their remarkably broad range of spin properties.[10,11] Graphene has been demonstrated as an excellent material for long distance spin transport due to the low spin-orbit coupling (SOC) and high electron mobility.[10,11,13,14] However, this low SOC remains a big challenge for the electrical control of the spin polarization in graphene.[11] In contrast, transition metal di-chalcogenides (TMDCs) are semiconductors with a high SOC, several *10 meV* and *100 meV* in the conduction and valence bands, respectively, which is orders of magnitude higher than in graphene.[10, 15] However, the very strong SOC hinders the realization of electrical spin transport in TMDCs themselves at room temperature.[16, 17]

To harvest these novel and contrasting spintronic properties of both 2D materials, hybrid devices, consisting of graphene and TMDC van der Waals heterostructures (vdWh)[18], are promising.[19] So far, the proximity induced strong SOC of graphene/TMDC vdWhs led to the observation of weak anti-localization[20] and spin Hall effect[21]. In addition to the advantage of spin transport in graphene, SOC in such vdWhs can be utilized to electrically control the spin polarization. A very recent report has shown the modulation of the spin signal at cryogenic temperatures by presenting spin valve measurements in such vdWhs.[22] However, for the active development in the field of spintronic and its possible future applications, realization of electrically tunable spin currents at ambient temperatures is a crucial requirement.

Here, we demonstrate the electronic control of the spin current in a spin field-effect transistor (spin-FET) device using graphene/TMDC vdWhs at room temperature. In particular, we fabricated heterostructures with large area chemical vapor deposited (CVD) graphene and TMDC molybdenum disulfide ($MoS_2$) flakes. The spin current is generated and detected by ferromagnetic (FM) tunnel contacts deposited on graphene (see Methods and Figure 1).[14] We demonstrate the spin-FET behavior by systematically investigating the electric gate control of both the non-local spin valve and Hanle spin precession signals. Our spin-FET satisfies four main fundamental requirements[23] at room temperature: 1. The electrical creation of spin current in graphene; 2. The diffusive transport of pure spin polarized electrons through the graphene/$MoS_2$ vdWh channel; 3. The non-local electrical detection of the spin current in spin valve and Hanle spin precession geometries; and 4. The crucial electrical control of the spin transport and life time by means of the gate voltage (Figure 1).

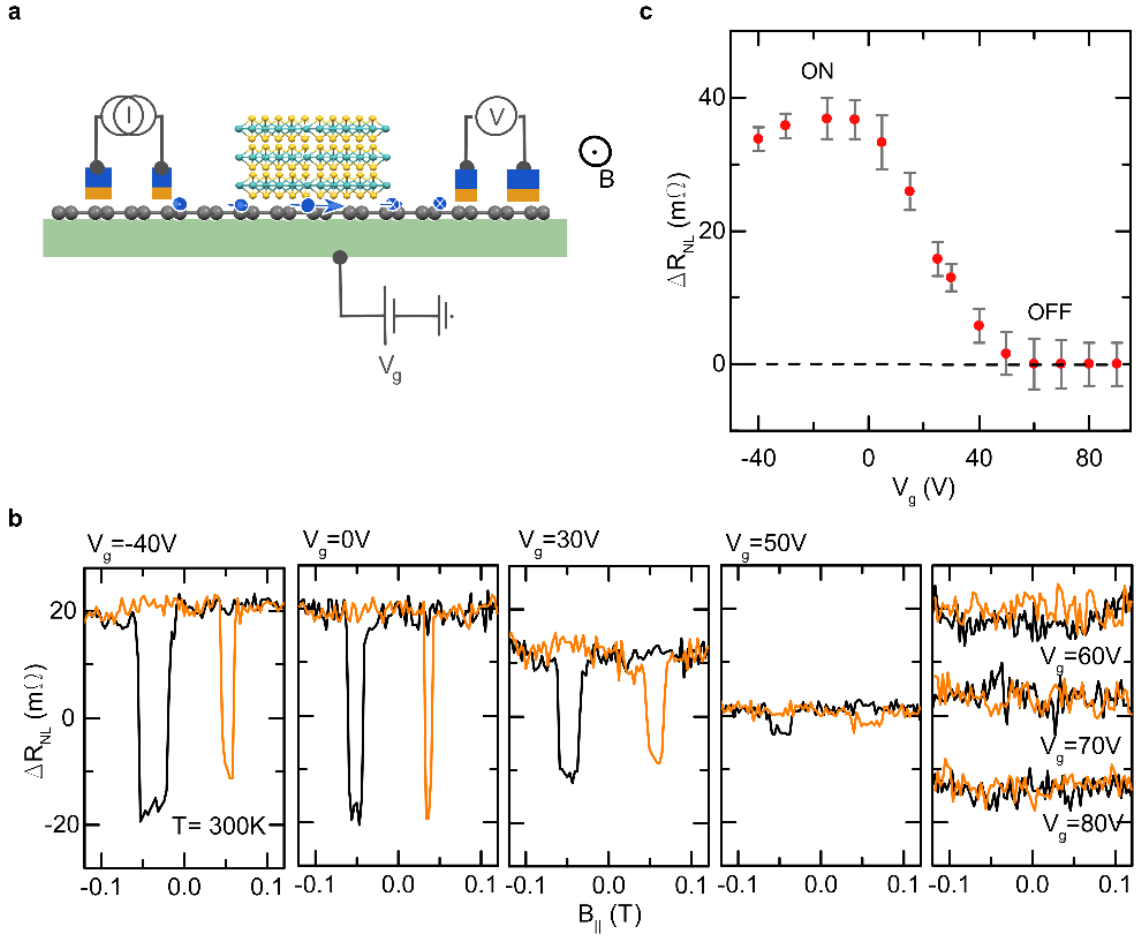

**Figure 2. Electrical gate control of the spin valve signal in graphene/MoS$_2$ heterostructures at room temperature. a.** Schematic of the spin–valve measurement geometry, where the spin current injector circuit (I) and the voltage detector circuit (V) are placed in a non-local (NL) geometry. The pure spin current diffusing in the heterostructure channel is detected as a voltage signal by the ferromagnetic detector. The magnetization of the injector/detector ferromagnetic contact and also the sign of spin accumulation are controlled by an in-plane magnetic field B$_\parallel$. **b.** NL spin valve magnetoresistance ($\Delta R_{NL} = \Delta V/I$) measurements at *300 K* by application of different gate voltages, *V$_g$*. Measurements are performed at a constant current source of *I= 30 µA*. A NL linear background is subtracted from the signal. **c.** spin-FET operation showing the dependence of the *ΔR$_{NL}$* on *V$_g$* at *300 K*.

To investigate the spin transport behavior in the graphene/MoS$_2$ heterostructure channel, spin valve measurements were performed in a non-local (NL) geometry at room temperature. Sweeping an in-plane magnetic field, the magnetization configuration of the injector and detector contacts (Figure 2a) can be aligned parallel or antiparallel, resulting in a magnetoresistance *ΔR$_{NL}$*. We observed a clear spin valve signal (Figure 2b) at room temperature, demonstrating the possibility of spin transport in the graphene/MoS$_2$ heterostructure with a channel length of *2.6μm* and a width of *0.6μm*. The NL spin valve signal $\Delta R_{NL} \approx 0.04\Omega$ is smaller than usually measured in pristine CVD graphene channels of similar geometry.[14] This reduction could arise from SOC induced in the graphene channel by MoS$_2$.[19,20,21]

Next, we measured the spin valve signal at different gate voltages (*V$_g$*) at room temperature (Figure 2b and Figure 2c). Below a threshold voltage *V$_g$<0V*, we observed spin valve signals *ΔR$_{NL}$* with almost constant amplitude. For the gate voltage range of 0<V$_g$<50V, a drastic reduction in *ΔR$_{NL}$* is observed, whereas for *V$_g$>50V* no spin signal could be detected. The strong modulation and vanishing spin current from ON to OFF state with application of gate voltage in our MoS$_2$/graphene heterostructure indicates a spin-FET operation at room temperature. This behavior in spin signal is reproducibly measured over different gate voltage sweep directions and also at low temperature (see Supplementary Information Figure S2 and Figure S3).

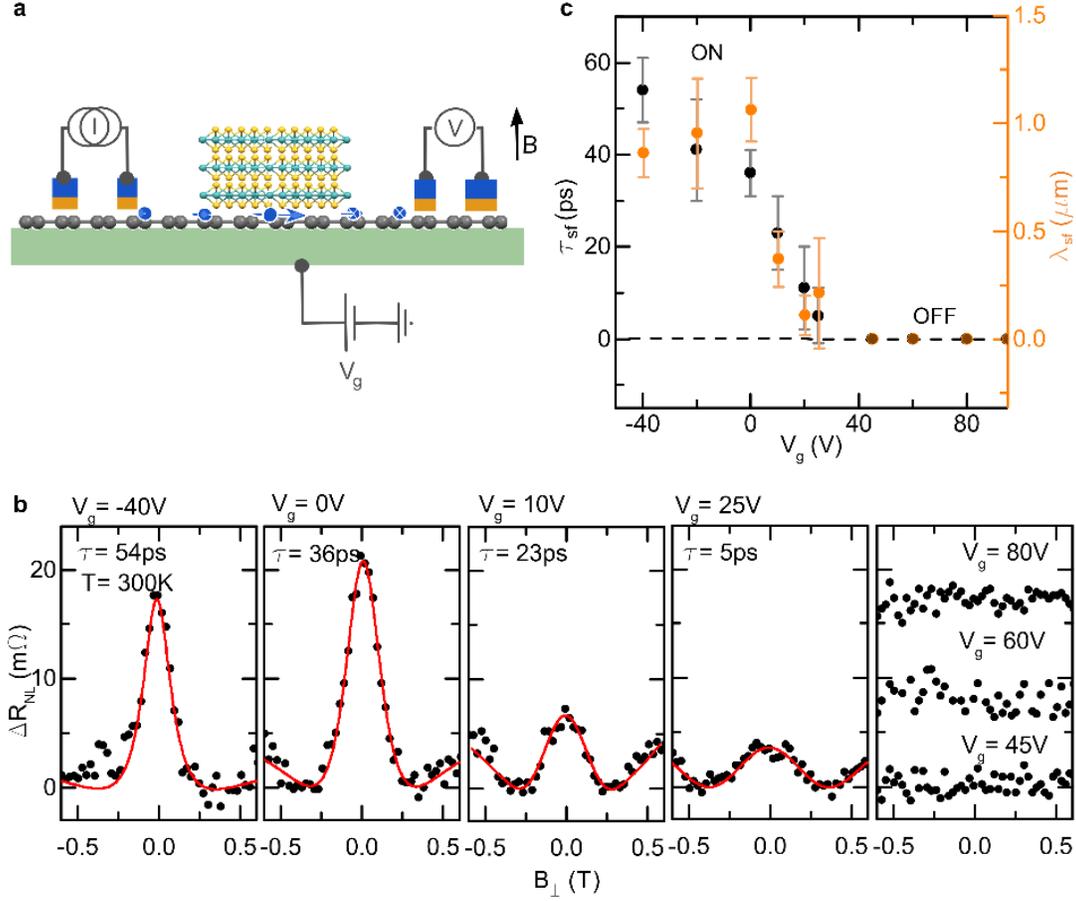

**Figure 3. Electrical gate control of the Hanle spin precession signal in graphene/MoS₂ heterostructures at room temperature**. **a.** NL Hanle geometry, where source and drain magnetization are aligned parallel while sweeping a perpendicular magnetic field $B_\perp$. **b.** NL Hanle spin signal $\Delta R_{NL}$ measured at *300 K* at different gate voltages. Measurements are performed at a constant current source of *I= 30 μA*. The raw data points are fitted with Eq. (1) (red line) to extract spin lifetime *τ$_{sf}$* and diffusion length $\lambda_{sf}$. **c.** The gate voltage V$_g$ dependence of spin lifetime *τ$_{sf}$* and diffusion length $\lambda_{sf}$ at *300 K*.

To confirm the spin-FET operation and understand the spin relaxation in the graphene/MoS₂ vdWhs, we performed NL Hanle spin precession measurements (Figure 3a). Figure 3b shows the Hanle data at room temperature for different gate voltages. In this geometry, the modulation of the signal stems from the spin precession about a perpendicular magnetic field (B$_\perp$) with the Larmor frequency $\omega_L = \frac{g\mu_B}{\hbar} B_\perp$ (Landé factor

g=2).[14,13] The variation of this non-local resistance $\Delta R_{NL}$ due to precession and relaxation of the spins diffusing from the injector to the detector can be described by

$$\Delta R_{NL} \propto \int_0^\infty \frac{1}{\sqrt{4\pi Dt}}\, e^{-\frac{L^2}{4Dt}}\, \cos(\omega_L t)\, e^{-\left(\frac{t}{\tau_{sf}}\right)} dt. \qquad (1)$$

With the channel length L, we can extract the spin lifetime $\tau_{sf}$ and diffusion constant $D_s \approx 0.006 - 0.025\, m^2 s^{-1}$ to calculate the spin diffusion length $\lambda_{sf} = \sqrt{D\tau_{sf}}$ (Figure 3c). Similar to the spin valve measurements, we observed an almost constant Hanle signal for $V_g<0V$ (Figure 3b). In this range, we obtain a spin lifetime $\tau_{sf} \approx 40 - 50\, ps$ and diffusion length $\lambda_{sf} \approx 0.8 - 1\, \mu m$. The reduced spin lifetime in our vdWhs, compared to pristine graphene devices (*200-300 ps*),[14] indicates a proximity induced SOC in graphene through $MoS_2$. Such a shortened spin lifetime has also been expected using the spin-orbit relaxation time $\tau_{so}$ determined from spin Hall effect studies[21], weak anti-localization measurements[20] and theoretical predictions[19]. In the gate voltage range of *0V<V_g<25V*, a strong tuning of the spin signal amplitude *ΔR_NL*, spin lifetime $\tau_{sf}$ and consequently diffusion length $\lambda_{sf}$ in the $MoS_2$/graphene channel is observed. The $\tau_{sf}$ can be tuned down to ~5 ps near *V_g = 25 V*. For *V_g>25V* the Hanle signal cannot be observed, validating a transition from ON to OFF state with gate voltage. The vanishing of the Hanle signal already at a gate voltage of about 30V is due to the reduction of the amplitude by factor two compared to the spin valve signal, which is consistent with the measurement principle (see Supplementary information Figure S3).[13] This confirms a spin-FET operation in our vdWhs at room temperature and reveals the tuning of the spin transport parameters in the channel. Our previous studies reported a weak modulation of up to 10 - 20 % in $\Delta R_{NL}$ and $\tau_{sf}$ by changing the carrier density in pristine graphene devices[14]. In our vdWhs, the observed strong reduction and vanishing of $\Delta R_{NL}$ and $\tau_{sf}$ is a clear characteristic of a spin-FET. Such a behavior in vdWhs could originate from a gate controlled interaction of spin polarized electrons in graphene with high SOC MoS$_2$.

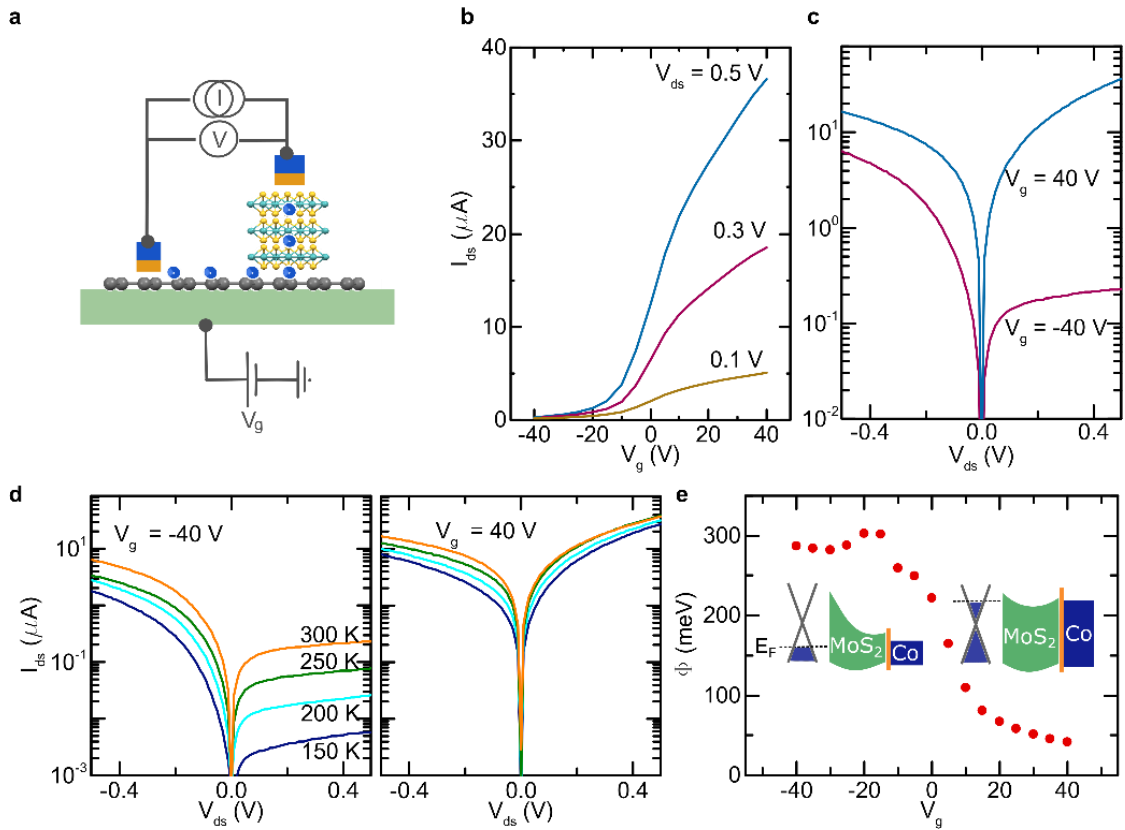

**Figure 4**: **Tuning the Schottky barrier at the MoS$_2$/graphene interface by gate voltage**. **a.** Schematics of the graphene/MoS$_2$ vdWh vertical device. **b.** Transfer characteristic (drain-source current I$_{ds}$ vs. gate voltage V$_g$ for different drain-source voltages V$_{ds}$) in a vertical device with MoS$_2$ thickness of 35nm. **c.** Output characteristics (I$_{ds}$ vs. V$_{ds}$) at $V_g = \pm 40V$. **d.** Output characteristics measured at different temperatures shown for V$_g$ = + 40 V (left panel) and V$_g$ = - 40 V (right panel). **e.** Schottky barrier height Φ obtained for different V$_g$. Inset: Band structures at the MoS$_2$/graphene interface for V$_g$ <0 and V$_g$ >0.

To understand the principle of our spin-FET operation we characterized the charge transport in both the MoS$_2$ channel and the graphene/MoS$_2$ vdWhs. The lateral MoS$_2$ FET shows a typical n-type transfer characteristic with an enhancement of current up to 10$^6$ times with application of V$_g$ at room temperature (see Supplementary Figure S4).

Such a gate-tunable MoS$_2$ channel resistance in our MoS$_2$/graphene vdWh spin-FET would allow the tuning of a parallel spin transport channel next to graphene. If the spin current from graphene can enter MoS$_2$ in the ON-state, spins would lose their coherence at a much faster rate, due to the high SOC of MoS$_2$.

Next, we investigated the charge transport in vertical graphene/MoS$_2$ devices, to understand the transport mechanisms at the interface (Figure 4a). The transfer characteristic of such a device is typical for n-type FETs with a current enhancement of more than 10$^2$ times (ON/OFF) at room temperature (Figure 4b). The corresponding output characteristic shows an asymmetric line shape in the OFF-state, which becomes symmetric in the ON-state (Figure 4c). This transistor behavior can be attributed to the presence of a gate-tunable Schottky barrier (Φ) at the MoS$_2$/graphene interface.[24] We determined this barrier Φ from the temperature dependent output characteristic (Figure 4d) using the thermionic emission model[16,24] (see Supplementary information Figure S5). Figure 4e shows the gate dependence of Φ at the graphene/MoS$_2$ interface, which changes from 300meV to 50meV when tuning the vertical transistor from OFF- to ON-state. This corresponds to a gate modulation of the graphene Fermi-level with respect to the MoS$_2$ conduction band as depicted in the insets of Figure 4e. Consequentially, in the spin-FET ON-state, a high Schottky barrier at the graphene/MoS$_2$ interface and high channel resistance of MoS$_2$ prevents spins in the graphene to enter the MoS$_2$ channel. In contrast, in the OFF-state of the spin-FET, the Schottky barrier at the graphene/MoS$_2$ interface is significantly reduced yielding an almost Ohmic contact, in addition to a drastic reduction in MoS$_2$ channel resistance. Under these conditions, spins can easily enter MoS$_2$, where they experience a much faster spin relaxation. This explains the reduction in spin lifetime in addition to the decrease in spin signal amplitude with gate voltage in our spin-FET device. This demonstrates that the gate voltage can be used to tune the band offset between material having complementary spintronics properties, facilitating the spins in graphene to interact efficiently with MoS$_2$.

The realization of a spin-FET at room temperature using 2D materials heterostructures is a significant step in the field of spintronics. Combining the gate dependence studies of spin valve, Hanle spin precession and Schottky barrier control in graphene/MoS$_2$ heterostructures, we demonstrate the spin-FET device operation. Utilizing the complementary spintronic properties of graphene and MoS$_2$ creates a unique platform, where the spin current can be tuned in a controlled manner by gate voltage. This graphene/MoS$_2$ spin-FET device scheme can be further improved by introducing insulating hexagonal boron nitride (h-BN) as tunnel barrier for efficient spin injection,[25,26] as substrate for improved graphene properties,[27,28] and as ultra-thin gate dielectrics for efficient gate control of the spin current in the channel. Furthermore, 2D semiconducting TMDCs of both electron and hole doped materials[29] can be combined with graphene to create complementary spin-FET devices for spin-logic operations. Beyond TMDCs, other emerging materials with topological protection could be used in heterostructures with graphene to create novel spin phenomena and provide opportunities for new discoveries.[30]

**Acknowledgements** We acknowledge financial supports from EU Graphene Flagship ( No. 604391), EU FlagEra projects (No. 2015-06813), Swedish Research Council young researcher grant (No. 2012-04604), Graphene center and the AoA Nano program at Chalmers University of Technology.

**Methods**

The van der Waals heterostructures were prepared using CVD graphene (Graphenea) on highly doped Si (with a thermally grown 285-nm-thick SiO$_2$ layer) and cleaned by Ar/H$_2$ annealing at 450ºC. The MoS$_2$ flakes (single crystals from SPI supplies) were transferred on top of graphene, which was patterned into individual channels either before or after the MoS$_2$ transfer. The graphene was patterned by photo- or electron beam lithography and oxygen plasma etching. Next, appropriate MoS$_2$ flakes of 10-35 nm located on graphene were identified by an optical microscope for device fabrication. The contacts were patterned on graphene (and MoS$_2$ flakes, in case of the vertical devices) by electron beam lithography. Finally, we used electron beam evaporation to deposit 8Å Ti, followed by *in situ* oxidation in a pure oxygen atmosphere for 30 min, resulting in an about 1nm thick TiO$_2$ layer. Without interrupting the vacuum, we deposited 80nm Co and finalized the devices by lift off in warm acetone at 60°C. In the final devices, the Co/TiO$_2$ contacts on graphene act as source and drain for spin polarized electrons, the MoS$_2$/graphene heterostructure region used as the channel, and the Si/SiO$_2$ is used as a gate for manipulation of the spin polarization. A Dirac curve for the graphene with MoS$_2$ on top is shown in Supplementary Figure S6 with charge neutrality point around V$_g$ of 16V. The measurements were performed in a cryostat with variable temperature and magnetic field facility. The current is applied using a Keithley 6221 current source and the non-local voltage is detected by a Keithley 2182A nanovoltmeter; the gate voltage was applied using a Keithley 2612 source meter.